\documentclass[12pt]{iopart}
\usepackage{amssymb}
\usepackage{graphicx,psfig,epsfig}
\usepackage{color}

\newcommand{\dd}{ {\textrm d}}              

\begin{document}

\title[Where Does the Energy Loss Lose Strength?]{Where Does the Energy Loss Lose Strength?}

\author{G G Barnaf\"oldi$^{12}$ , G Fai$^{1}$, P L\'evai$^{2}$, G Papp$^{3}$ , B A Cole$^{4}$ }

\address{$^{1}$\ CNR, Kent State University, Kent, OH 44242, USA} 
\address{$^{2}$\ RMKI KFKI, P.O. Box 49, Budapest, H-1525, Hungary} 
\address{$^{3}$\ E\"otv\"os University, P\'azm\'any P. S\'et\'any 1/A, 
Budapest, H-1117, Hungary}
\address{$^{4}$\  Nevis Laboratory, Columbia University, New York, NY, USA}

\ead{bgergely@rmki.kfki.hu}

\begin{abstract}
Nuclear modification factors for pion production in $AuAu$ and 
$CuCu$ collisions are analyzed at very high transverse momenta. 
At $p_T \gtrsim 10$ GeV/c, the $R_{AA}(p_T)$ is determined mostly by
the initial state nuclear modifications (e.g. EMC effect) and the 
non-Abelian jet-energy loss in the final state. At high momenta these 
effects together are strong enough to suppress $R_{AA}(p_T)$ to below 
$1$ at RHIC energies. We display results using HKN shadowing in our 
pQCD improved parton model. Result of a similar calculation at LHC 
energies for $PbPb$ collisions are also displayed. Based on 
$\dd N/\dd y$ estimates, a larger opacity value, 
$L/\lambda_g\approx 10 \pm 2$, is used for the produced partonic 
matter in central collisions at the LHC. 
\end{abstract}
\submitto{\JPG}


\section*{Introduction}

Suppression of the inclusive pion spectra in central $AuAu$ collisions 
at RHIC relative to the peripheral (or $pp$) spectra can be understood
in terms of non-Abelian energy loss~\cite{BDMPS,GLV} in the quark-gluon 
plasma (QGP) created in central nuclear 
collisions~\cite{phenix_QM01,LP_QM01}. Early PHENIX data on the nuclear 
modification factor, $R_{AuAu}(p_T)$ explored the transverse momentum 
range 2 GeV/c $\lesssim p_T \lesssim$ 10 GeV/c, where the medium-induced 
non-Abelian relative energy loss was found to be almost constant as a 
function of $p_T$. A factor of $\sim 5$ suppression was 
measured~\cite{bias2_phenix,bias_phenix,ismd}, and a fractional energy 
loss, $S_{loss}$, was extracted~\cite{azimut_phenix}.  

The most recent (preliminary) high-$p_T$ data on 
$R_{AA}(p_T)$~\cite{new_phenix} in central and mid-central $CuCu$ and 
$AuAu$ collisions cover the $p_T$ range up to $\approx 20$ GeV/c, and 
show a slight increase with $p_T$. Such a decrease of the strength of the 
energy loss is expected at very high transverse momenta, even though 
the EMC suppression is also present in this region\cite{Cole:2007ru}.   

In this paper we study the Gyulassy-L\'evai-Vitev (GLV)~\cite{GLV} energy 
loss at high transverse momenta in central $AA$ collisions at RHIC and 
LHC energies. We concentrate on the $p_T\gtrsim 5$ GeV/c region, where 
pQCD results can be trusted. To check the scaling properties, we present 
$R_{AuAu}$ as a function of both, $p_T$ and $x_T$. We expect the nuclear 
modifications to scale with $x_T$ and to lose strength at high opacity 
and large transverse momenta.  

\section*{Scaling of GLV Jet Quenching}

The slight increasing trend in the data on $R_{AA}(p_T)$ in 
Ref~\cite{new_phenix} beyond $p_T\gtrsim 10$ GeV/c both in $AuAu$ and 
$CuCu$ collisions should be analyzed based on the high-$p_T$ behavior 
of nuclear shadowing and energy loss. The medium-induced non-Abelian 
energy loss, which is expected to be the stronger of the two effects, 
is described by    
\begin{equation}
\Delta E \approx \frac{C_R \alpha_s}{N(E)} \cdot \frac{L^2 \mu^2}{\lambda_g} 
\cdot \log \left[ \frac{E}{\mu}\right] \approx \frac{C_R \alpha_s}{N(E)} 
\cdot \frac{1}{A_{\perp}}\frac{\dd N}{\dd y} \cdot \langle L \rangle 
\cdot \log \left[ \frac{E}{\langle \mu \rangle }\right] \,\, ,
\end{equation}
where the meaning of the various symbols can be found in the GLV 
reference\cite{GLV}. Most important for the present study is the fact 
that the implied relative energy loss, $\Delta E/E$, has a maximum at 
around $p_T\sim 5$ GeV/c (which depends on the parameters weakly), and 
$\Delta E/E$ is almost constant in the $2$ GeV/c 
$\lesssim p_T \lesssim 10$ GeV/c energy region for opacity values 
$L/\lambda_g \lesssim 2-4$ (see Ref.~\cite{LP_QM01}).     

\begin{figure}[ht]
\begin{center} 
\resizebox{75mm}{50mm}{\includegraphics{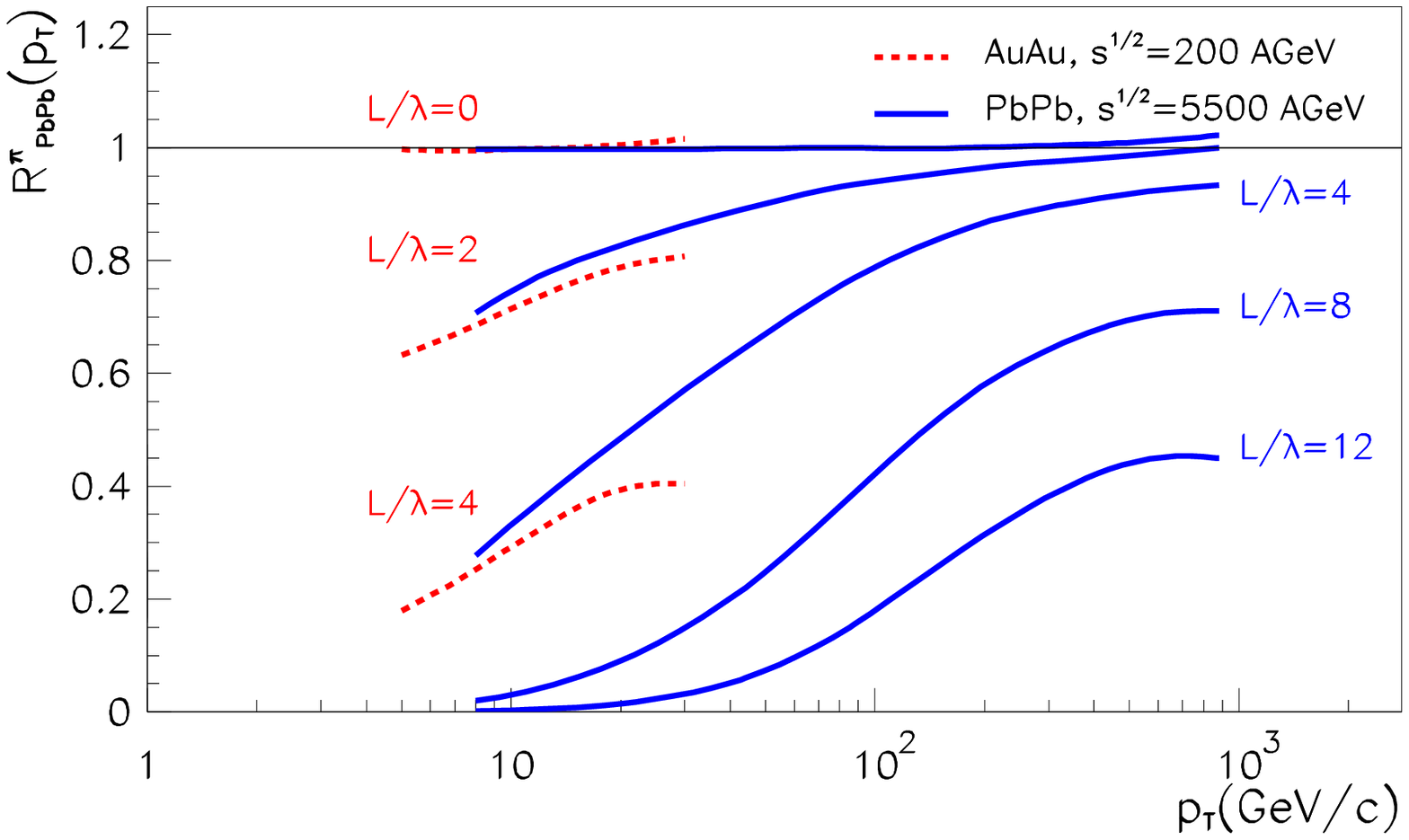}}
\resizebox{75mm}{50mm}{\includegraphics{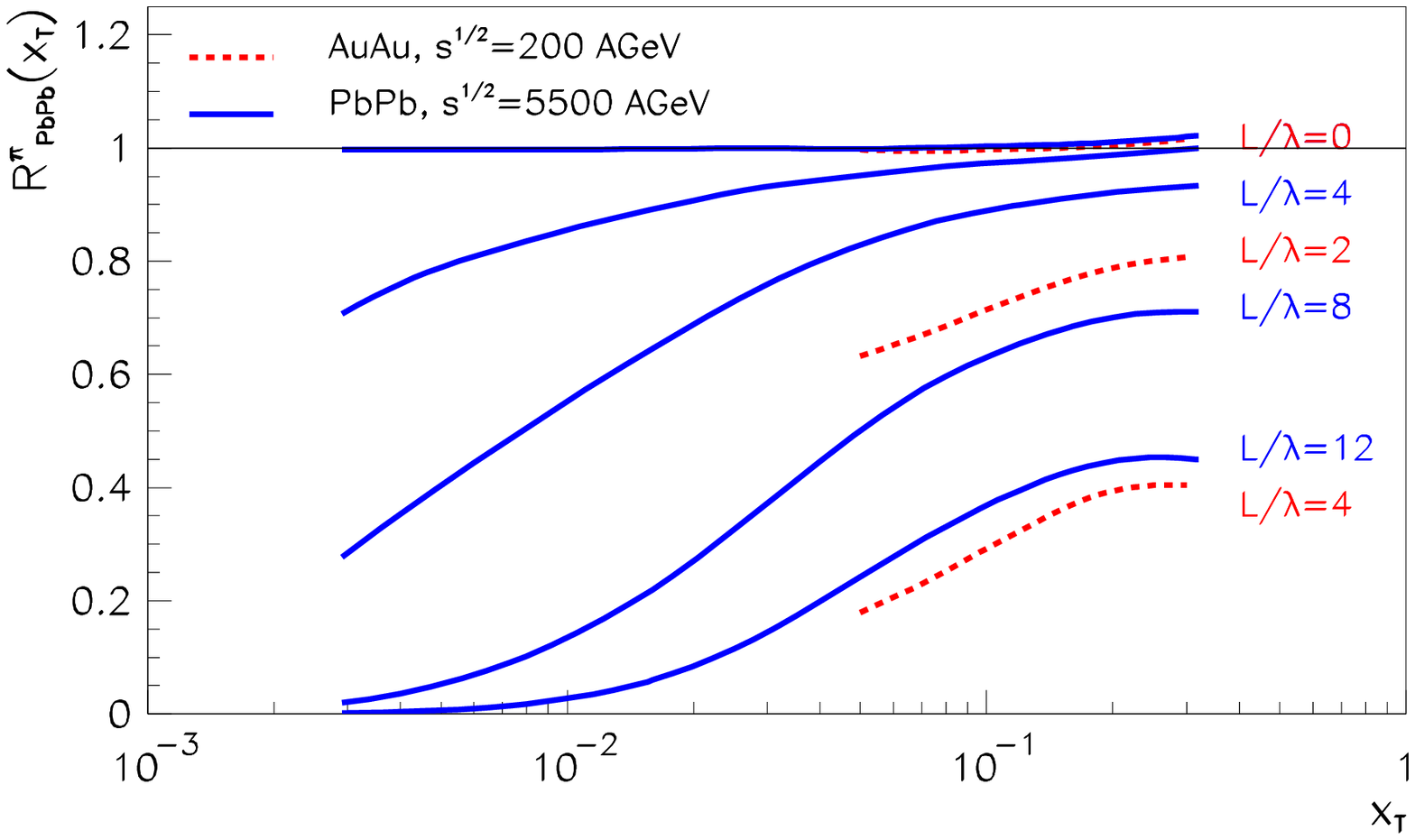}}
\caption{Scaling of energy loss effects in most central $AuAu$ 
and $PbPb$ collisions with $p_T$ ({\sl left panel}) and with $x_T$ 
({\sl right panel}).}
\label{fig:2}
\end{center}
\end{figure}

At higher opacities the maximum is relatively stable at around 
$E\sim 4-5$ GeV, and it becomes more well-defined with increasing opacity.
The effect of jet quenching thus becomes very strong around its maximum.
At the same time, the energy loss is no longer a constant, but becomes  
an $E$-dependent quantity. Due to the $\sim \log{(E/\mu)}$ tail of the 
distribution, the energy loss of highly energetic jets is getting weaker, 
thus less suppression is expected at high $p_T$ values. Theoretical 
expectation for $\dd N /\dd y$ give $1200 \pm 300 $ for the most 
central $AuAu$ collisions at RHIC\cite{GLV,LP_QM01,new_phenix}, 
corresponding 
to $L/\lambda_g \approx 4$, while  $\dd N /\dd y \sim (1500-4000)$ 
is a rough estimate for LHC energies, making $L/\lambda_g$ up to 12 
relevant for calculations at LHC energies~\cite{lastcall}. 

To see the effect of the energy loss clearly, we present in 
Fig.~\ref{fig:2} the generated nuclear modification with all initial 
and other final state nuclear effects switched off, using the above 
values. The left panel shows $R^{\pi}_{AA}(p_T)$, while in the right 
panel we display $R^{\pi}_{AA}(x_T)$. 

With opacity $L/\lambda_g=0$, the nuclear modification is naturally 
$\sim 1$. The strongest suppression is associated with the highest 
density of the medium, reflected here in $L/\lambda_g=8-12$. 
It should be kept in mind that, due to the different energy densities 
at RHIC and LHC, a larger opacity value at LHC is expected to lead to 
a suppression similar to the one obtained at RHIC with a smaller 
$L/\lambda_g$. Comparing the left and right panels of Fig.~\ref{fig:2}, 
we conclude that $x_T$-scaling of $R_{AA}$ from RHIC (dashed lines) to 
LHC (solid lines) energies is more adequate than $p_T$ scaling. 
\begin{figure}[ht]
\begin{center}
\resizebox{75mm}{50mm}{\includegraphics{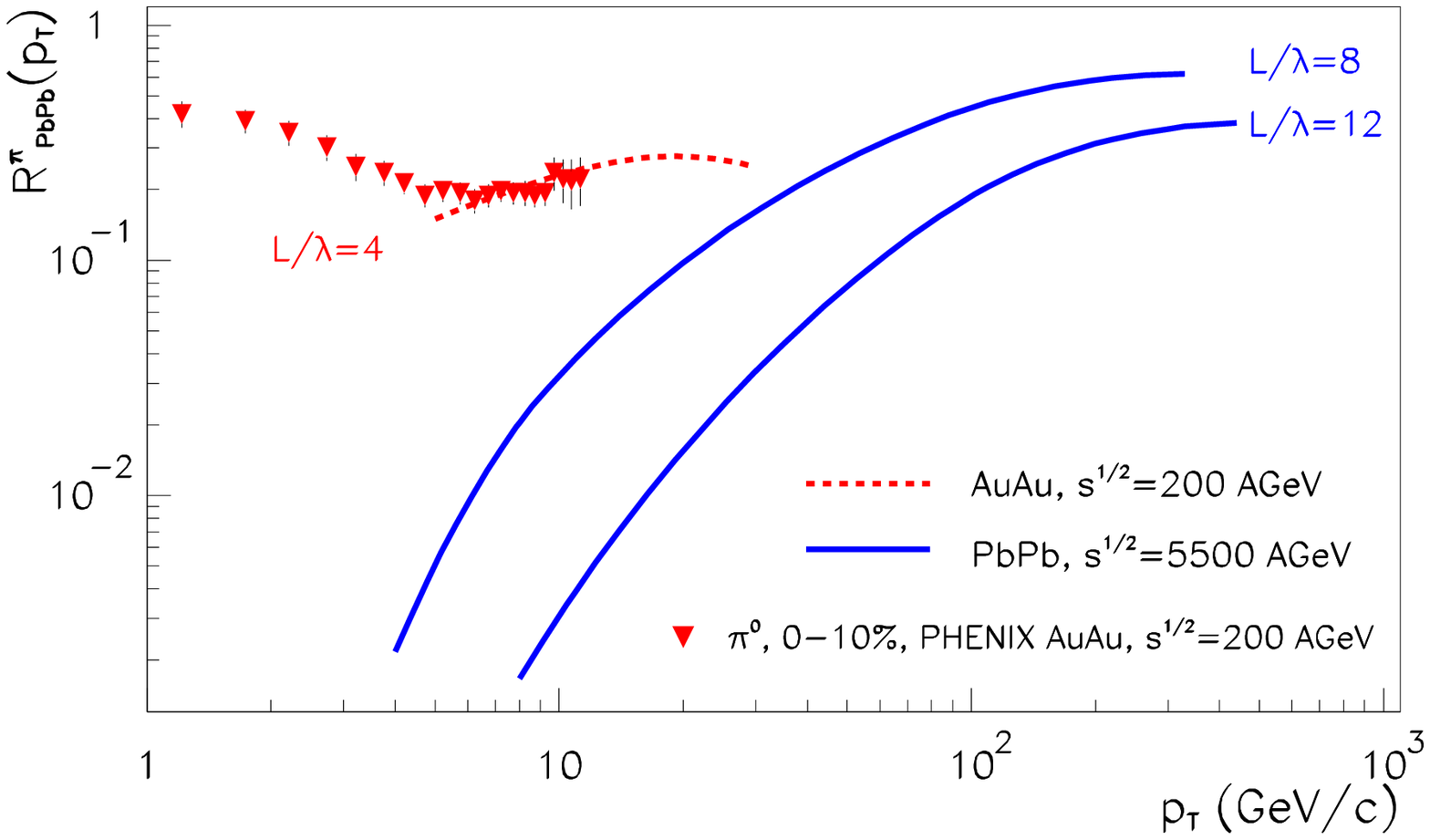}}
\resizebox{75mm}{50mm}{\includegraphics{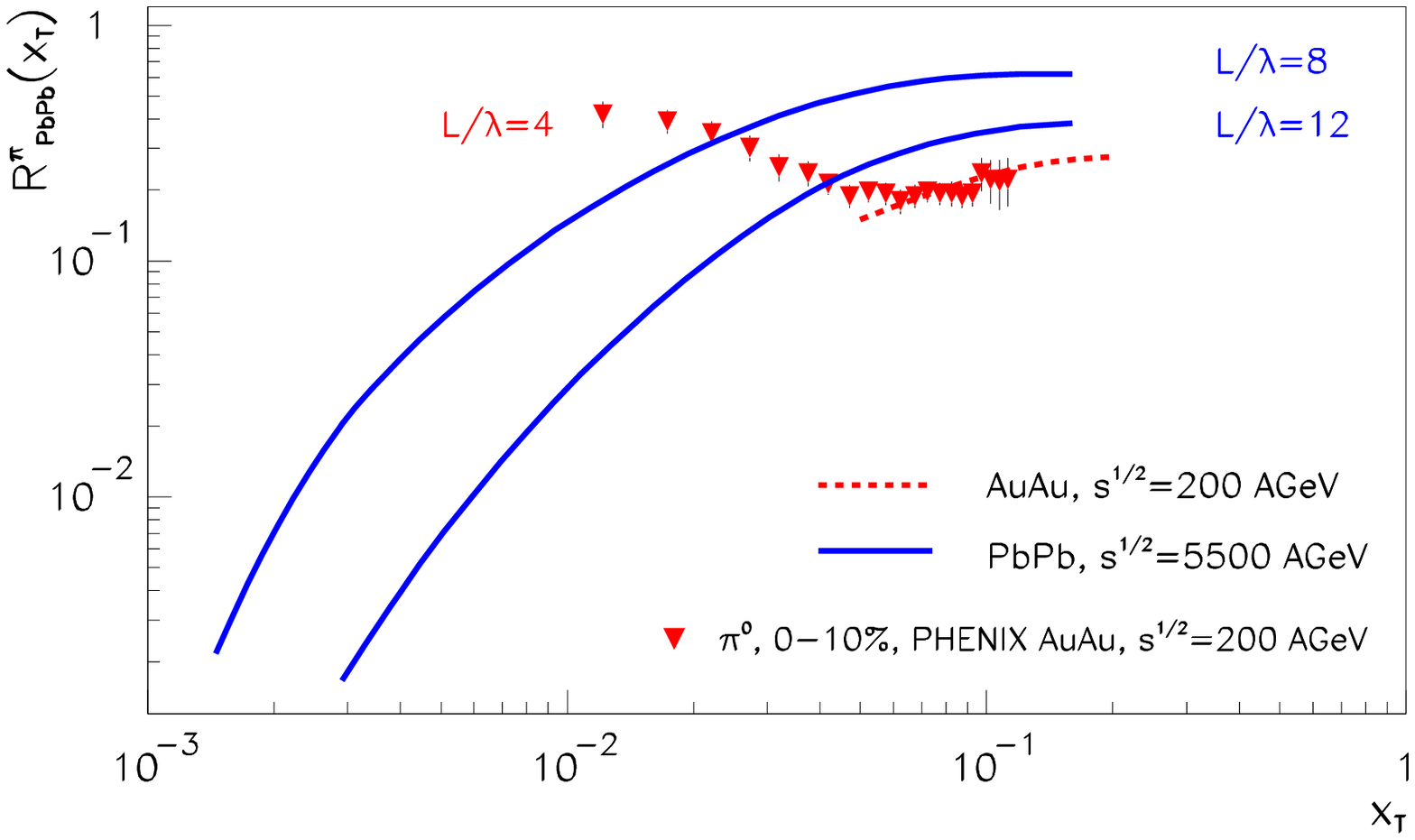}}
\caption{Calculations and data for nuclear modifications in most 
central $AuAu$ and $PbPb$ collisions including final and initial 
state effects. (See text for details.)}
\label{fig:3}
\end{center}
\end{figure}

Fig.~\ref{fig:3} includes the initial state nuclear effects, using 
HKN~\cite{HKN} shadowing in the pQCD improved parton model~\cite{Yi02}. 
(The notation is similar to that of Fig.~\ref{fig:2}.) It can be seen 
from the left panel that after a minimum, the data on $R_{AA}$ 
increase toward high $p_T$, due to the logarithmic tail of the energy 
loss. At $p_T \gtrsim 10$ GeV/c the calculated nuclear modification factor 
exhibits a maximum and then decreases again, reaching the EMC
region. As seen on the right panel, the convolution with the shadowing 
function preserves the $x_T$ scaling of the nuclear 
modification~\cite{GGB_Poster}. Based on the scaling, we attribute an 
approximate opacity parameter $\sim 10\pm 2$ to central $PbPb$ 
collisions at $\sqrt{s}=5.5$ $A$TeV LHC energy. Note, however,
that in minimum bias measurements an averaged opacity should be 
relevant, much smaller than the central maximum value. It should also 
be remembered that the suppression has an angular dependence, which 
needs to be tested by further geometrical studies of jets.

\section*{Concluding Remarks}

We studied the non-Abelian jet energy loss in the first-order 
GLV framework. Concentrating on the perturbative, high-$p_T$ region, we
pointed out that the maximum of the energy loss becomes sharper, and 
that the phenomenon loses strength due to the logarithmic tail at 
high $p_T$ and opacities greater than $4$. At larger opacities the 
standard picture of a constant relative energy loss needs to be modified. 
Thus, energy-loss effects are more complicated than a simple shift of the 
high-$p_T$ spectra. This may explain why $S_{loss}$ does not show 
perfect scaling for different system sizes or 
centralities~\cite{new_phenix}.  

We found that the energy loss scales with $x_T$ (using appropriately 
different opacity values), and the 
introduction of initial-state nuclear effects preserves this scaling. 
Based on Fig.~\ref{fig:2} we extract the opacity parameter 
$\sim 10 \pm 2$ for most central $PbPb$ collisions at $\sqrt{s}=5.5$ 
$A$TeV LHC energy.

The extracted $L/\lambda_g$ values are high, but should only appear in 
the most dense region of the collision. The present description of the 
jet is incomplete, since the geometrical properties of the jet were 
not taken into account. A detailed analysis should investigate the 
angular size variation of the jet and the angular distribution of 
the energy loss inside the jet.   

\section*{Acknowledgments}
One of the authors (GGB) would like to thank to the organizers for local 
support. Our work was supported in part by Hungarian OTKA PD73596, 
T047050, NK62044, and IN71374, by the U.S. Department of Energy under 
grant U.S. DE-FG02-86ER40251, and jointly by the U.S. and Hungary under 
MTA-NSF-OTKA OISE-0435701.


%
\end{document}